\documentclass{jfm}

\usepackage{natbib}

\newcommand{\refeq}[1]{(\ref{#1})}

\newcommand{\be}{{\bf e}}

\newcommand{\br}{{\bf r}}

\newcommand{\bv}{{\bf v}}

\newcommand{\bnabla}{\mbox{\boldmath$\nabla$}}

\newcommand{\diff}{{\,\mathrm d}}

\newcommand{\smallfrac}[2]{{\textstyle\frac{#1}{#2}}}
\newcommand{\halff}{{\textstyle\frac{1}{2}}}

\newcommand{\quarter}{{\textstyle\frac{1}{4}}}

\newcommand{\const}{\mbox{const}}

\newlength{\tightsecu}
\newlength{\tightsecl}
\newlength{\tightsubsecu}
\newlength{\tightsubsecl}
\newlength{\tightsubsubsecu}
\newlength{\tightsubsubsecl}
\newlength{\sectosubsec}
\newlength{\subsectosubsubsec}
\newcommand{\lb}{\left (}
\newcommand{\rb}{\right )}

\newcommand{\eFlow}{\hat\be_\flowCoordinate}

\newcommand{\crit}{{\rm crit}}

\newcommand{\flowCoordinate}{x}
\newcommand{\vorticityCoordinate}{y}
\newcommand{\gradientCoordinate}{z}

\newcommand{\ez}{\hat{\be}_z}

\newcommand{\relativePosition}{\Delta\mathbf{r}}
\newcommand{\xOffset}{\Delta x}
\newcommand{\yOffset}{\Delta y}
\newcommand{\zOffset}{\Delta z}
\newcommand{\xOffsetCrit}{\Delta_\crit}

\newcommand{\relativeVelocityVector}{\Delta\mathbf{U}}
\newcommand{\relativeVelocity}{\Delta U}

\newcommand{\rhoOffset}{\Delta\rho}
\newcommand{\rhoOffsetVector}{\Delta\boldsymbol{\rho}}
\newcommand{\rhoOffsetCrit}{\Delta_\crit}

\newcommand{\gap}{\epsilon}
\newcommand{\gapMinSwap}{\epsilon_{\min}}

\newcommand{\xField}{x}

\newcommand{\zField}{\bar z}
\newcommand{\rField}{\bar r}
\newcommand{\rFieldVector}{\bar\br}
\newcommand{\rFieldUnitVector}{\hat{\bar\br}}

\newcommand{\xFieldR}{\xField}

\newcommand{\zFieldR}{\zField{'}}
\newcommand{\rFieldR}{\rField{'}}
\newcommand{\rFieldVectorR}{\rFieldVector{'}}
\newcommand{\rFieldUnitVectorR}{\rFieldUnitVector{'}}

\newcommand{\shearRate}{\dot{\gamma}}

\newcommand{\wallVelocity}{\mathbf{U}_{\rm w}}
\newcommand{\externalFlow}{\mathbf{v}^{\rm ext}}
\newcommand{\scatteredFlow}{\delta\mathbf{v}_1}
\newcommand{\scatteredFlowComponent}[1]{\delta v_{1#1}}
\newcommand{\scatteredFlowComponentW}[1]{\delta v_{1#1}^*}

\newcommand{\viscosity}{\eta}

\newcommand{\chiGap}{\chi_\gap}
\newcommand{\chiSwap}{\chi_{\textrm{swap}}}

\newcommand{\self}{{\mathrm{s}}}
\newcommand{\swap}{{\mathrm{swap}}}
\newcommand{\rough}{{\mathrm{rough}}}
\newcommand{\selfDiffusivity}{D_{\self}}
\newcommand{\selfDiffusivitySwap}{D_{\self}^{\swap}}
\newcommand{\selfDiffusivitySwapND}{{\tilde D}_{\self}^{\swap}}
\newcommand{\selfDiffusivityRough}{D_{\self}^{\rough}}

\newcommand{\maxx}{{\textrm{max}}}
\newcommand{\zOffsetMax}{\zOffset_\maxx}

\newcommand{\yOffsetND}{\Delta\tilde y}
\newcommand{\zOffsetND}{\Delta\tilde z}
\newcommand{\zOffsetMaxND}{\Delta{\tilde z}_\maxx}

\newcommand{\roughnessParameter}{\epsilon_{\textrm{r}}}

\newcommand{\oP}{{\rm \hat P}}

\usepackage{amsmath}
\usepackage{amssymb}
\usepackage{graphicx}
\usepackage{amsbsy}

\usepackage{epsfig}

\title[New cross-streamline particle migration mechanism]
{Swapping trajectories: a new wall-induced cross-streamline particle
migration mechanism in a dilute suspension of spheres}
\author[M.Zurita-Gotor, J. B{\l}awzdziewicz and E. Wajnryb]
{
M.\ns Z\ls U\ls R\ls I\ls T\ls A\ls -\ls G\ls O\ls T\ls O\ls R,$^{1,2}$\break
J.\ns B\ls \L\ls A\ls W\ls Z\ls D\ls Z\ls I\ls E\ls W\ls I\ls C\ls Z,$^1$
\ns\and E.\ns W\ls A\ls J\ls N\ls R\ls Y\ls B$^3$
}
\affiliation{
$^1$Department of Mechanical Engineering, P.O. Box 20-8286, Yale
University, New Haven, CT 06520-8286, USA\\[\affilskip]
$^2$Present address: Departamento de Ingenieria Aeroespacial y
Mecanica de Fluidos, Universidad de Sevilla Camino de los
descubrimientos s/n Sevilla 41092, Spain\\[\affilskip]
$^3$IPPT, \'Swi\c{e}tokrzyska 21, Warsaw, Poland
}

\date{\today}

\begin{document}
\maketitle

\begin{abstract}
Binary encounters between spherical particles in shear flow are studied for a
system bounded by a single planar wall or two parallel planar walls under
creeping flow conditions.  We show that wall proximity gives rise to a new
class of binary trajectories resulting in cross-streamline migration of the
particles.  The spheres on these new trajectories do not pass each other (as
they would in free space) but instead they swap their cross-streamline
positions.  To determine the significance of the wall-induced particle
migration, we have evaluated the hydrodynamic self-diffusion coefficient
associated with a sequence of uncorrelated particle displacements due to
binary particle encounters.  The results of our calculations quantitatively
agree with the experimental value obtained by \cite{Zarraga-Leighton:2002} for
the self-diffusivity in a dilute suspension of spheres undergoing shear flow
in a Couette device.  We thus show that the wall-induced cross-streamline
particle migration is the source of the anomalously large self-diffusivity
revealed by their experiments.

\end{abstract}

\section{Introduction}
\label{Introduction}

Random displacements resulting from particle encounters in suspension flows
lead to hydrodynamically induced particle migration with respect to the local
suspension velocity
\citep{%
Eckstein-Bailey-Shapiro:1977,%
Leighton-Acrivos:1987,%
Bossis-Brady:1987%
}.
For non-Brownian particles under creeping-flow conditions, hydrodynamically
induced migration constitutes an important mechanism of particle
redistribution in the suspending fluid.  Thus studies of particle encounters
in flowing suspensions are vital both from the fundamental and practical
points of view.

As shown by \cite{Batchelor-Green:1972a}, two spheres passing each other in
unbounded shear flow return to the initial transverse (cross-streamline)
positions after a binary encounter is completed.  This behavior follows from
the flow reflection symmetry of Stokes equations and a reflection symmetry of
the system.  Since on open binary trajectories there are no cross-streamline
particle displacements, for perfect spheres in free space the transverse
hydrodynamic-diffusion process requires encounters of at least three particles
\cite*%
[see, eg.][]
{%
Acrivos-Batchelor-Hinch-Koch-Mauri:1992,%
Wang-Mauri-Acrivos:1996,%
Wang-Mauri-Acrivos:1998,%
Drazer-Koplik-Khusid-Acrivos:2002%
}.
It follows that the self-diffusion coefficient scales as $D_{\mathrm
s}\sim O(\phi^2)$ in the low-concentration regime, and the $O(\phi)$
contribution associated with binary encounters arises only in the
presence of non-hydrodynamic forces that remove the flow-reflection
symmetry of the problem.  For non-Brownian particles non-hydrodynamic
interactions often result from direct particle contacts due to
surface roughness. 

Several years ago \cite{Zarraga-Leighton:2002} measured the $O(\phi)$
contribution to the shear-induced self-diffusivity $\selfDiffusivity$
for a dilute suspension of spheres undergoing shear flow in a Couette
device.  The experiments yielded a surprising result: the
self-diffusion coefficient was nearly an order of magnitude higher
than the theoretical estimate
\citep{da_Cunha-Hinch:1996,Zarraga-Leighton:2001} for rough spheres
with the roughness amplitude corresponding to the experimental system.
Several possible causes of the anomalous self-diffusivity (such as
inertial lift and non-Newtonian effects) were examined, but none of
them was sufficient to explain the anomaly.  

The suspension in the Couette device used in experiments of
\cite{Zarraga-Leighton:2002} was bounded by nearly flat parallel walls
separated by a relatively small distance $H=20d$ (where $d$ is the
particle diameter).  In the analysis of experimental results it was
assumed that the effect of walls on the self-diffusivity coefficient
must be negligible.  Since the fore-aft symmetry of the system would
ensure that after passing each other the particles would return to
their original streamlines (as they do in infinite space), it seemed
unlikely that the walls were the cause of the enhanced
self-diffusivity.

In the present paper we show that hydrodynamic interactions of particle pairs
with the confining walls result in cross-streamline particle displacements in
binary encounters in shear flow.  Specifically, we have found a new class of
binary trajectories: the particles on such trajectories initially approach
each other, but then they move across the channel in opposite directions and
separate without passing each other (unlike the particles in free space).

The new class of trajectories results from a sign change of the transverse
component of the relative particle velocity.  Such sign changes in
wall-bounded systems were first pointed out in our recent
study \citep*{Bhattacharya-Blawzdziewicz-Wajnryb:2005}.  Thus seemingly subtle
features of particle mobilities produce a significant qualitative effect: our
explicit calculations show that after the new class of particle trajectories
is included in an estimate for the shear-induced self-diffusivity, a
quantitative agreement is obtained between the measurements of
\cite{Zarraga-Leighton:2002} and theoretical predictions.  Hence, the paradox
of the unusually large self-diffusivity observed by
\cite{Zarraga-Leighton:2002} has been resolved.

The system considered in our paper is defined in section \ref{system
definition}.  The new class of binary trajectories resulting in
cross-streamline particle displacements and the physical mechanism leading to
this behavior is discussed in section \ref{Effect of walls on binary collision
dynamics}.  In section \ref{mixing section} we analyze the consequences of the
new pair trajectories for cross-streamline particle migration in suspensions
of spheres in the low-concentration regime.  Our findings are summarized in
section \ref{Conclusions}.

\begin{figure}
  \centering
  \includegraphics[width=1\textwidth]{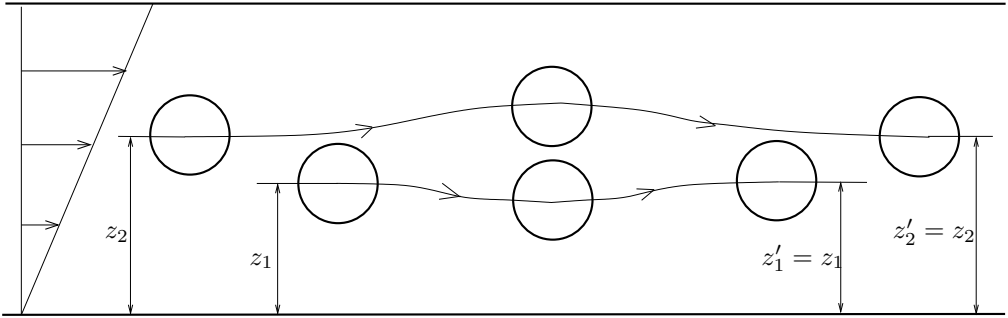}
\caption{
System definition and schematic representation of a non-swapping
binary trajectory.
}
\label{sketch}
\end{figure}

\section{Definition of the system}
\label{system definition}

We consider the dynamics of binary encounters of spherical particles
undergoing stationary shear flow in a space bounded by a single planar wall or
two parallel planar walls separated by the distance $H$.  We focus on
configurations where the particle--wall separation is comparable to the sphere
diameter $d$. The suspending fluid has viscosity $\viscosity$, and creeping
flow conditions are assumed.

We use a coordinate system where the walls are in the
$\flowCoordinate$--$\vorticityCoordinate$ planes.  The positions of particle
centers are denoted by $\mathbf{r}_i$ and the linear and angular velocities of
the particles by $\mathbf{U}_i$ and $\boldsymbol{\Omega}_i$, respectively,
where $i=1,2$.  The walls are at $\gradientCoordinate=0$ and
$\gradientCoordinate=H$. Particle encounters are described using the relative
coordinates $\relativePosition=(\xOffset,\yOffset,\zOffset)$, where
\begin{equation}
\label{relative position}
\relativePosition=\mathbf{r}_2-\mathbf{r}_1
\end{equation}
is the relative position vector centered on particle $1$.

The unperturbed fluid velocity
\begin{equation}
  \label{external velocity}
  \externalFlow \lb \mathbf{r} \rb = \shearRate \gradientCoordinate \eFlow
\end{equation}
(where $\shearRate$ denotes the shear rate) points in the lateral
direction $\flowCoordinate$, and varies in the normal direction
$\gradientCoordinate$. The flow occurs due to the motion of
the upper wall with velocity
\begin{equation}
  \label{wall velocity}
  \wallVelocity=\shearRate H \eFlow.
\end{equation}
Particles are torque and force free. No-slip boundary conditions are
imposed at both walls and at the particle surfaces.

In our calculations of particle trajectories, interparticle and
particle--wall hydrodynamic interactions are accurately taken into
account.  For a one-wall system the particle velocities are evaluated
using a reflection technique
\citep{Cichocki-Jones-Kutteh-Wajnryb:2000}, and for the two-wall
system we use the Cartesian-representation method, recently developed
by our group
\citep{%
  Bhattacharya-Blawzdziewicz-Wajnryb:2005a,%
  Bhattacharya-Blawzdziewicz-Wajnryb:2005,%
  Bhattacharya-Blawzdziewicz-Wajnryb:2006%
}.
The equations of motion $\mathbf{\dot{r}}_i=\mathbf{U}_i$ (where the
dot denotes the time derivative) are integrated using a Runge-Kutta
algorithm with an adaptive time step
\citep{Press-Vetterling-Flannery-Teukolsky:1992}.

\begin{figure}
  \centering
  \includegraphics[width=1\textwidth]{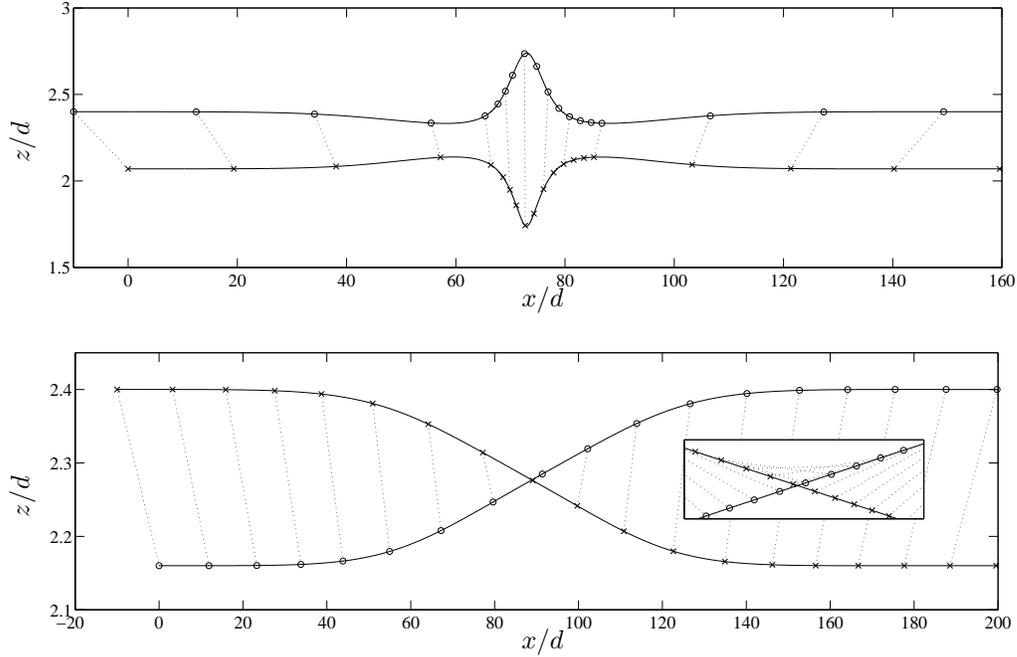}

\caption{
Examples of non-swapping (top panel) and swapping (bottom panel) trajectories
for two spheres in shear flow (\protect\ref{external velocity}) between
parallel planar walls.  The walls are at $z=0$ and $z=5d$, where $d$ is the
particle diameter.  The lower wall is at rest, and the spheres are moving in
the flow--gradient plane $x$--$z$.  Equal-time positions of sphere centers are
connected by dotted lines.  The inset shows the blowup of the
trajectory-intersection region.
}
\label{sample trajectories}
\end{figure}

The geometry of the system and a sketch of trajectories for two spheres that
pass each other are shown in figure \ref{sketch}.  The trajectories in this
figure are out of scale, and an accurate representation of the trajectories
can be seen in figure \ref{sample trajectories}.

\begin{figure}
  \centering
  \includegraphics[width=1\textwidth]{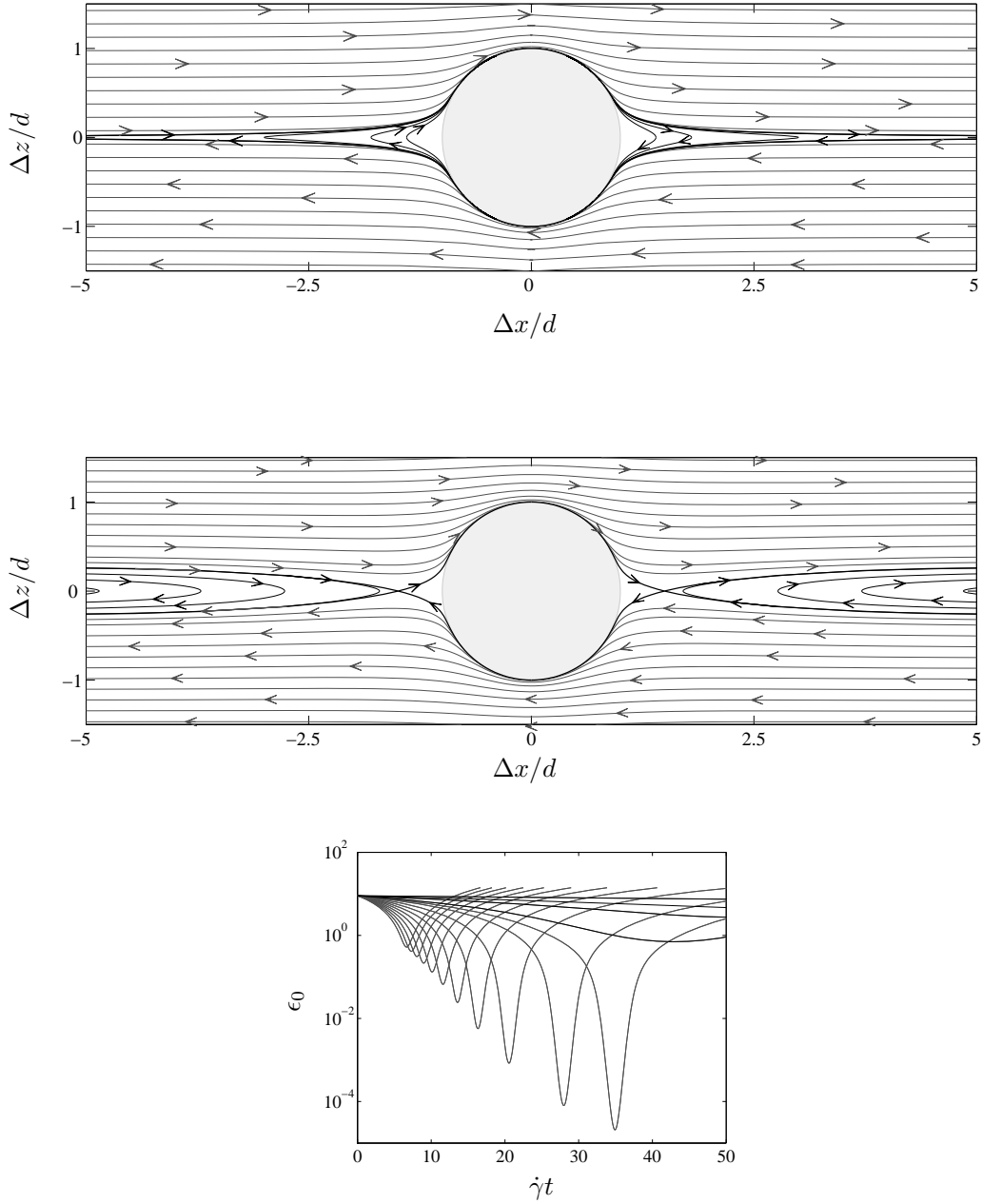}
\caption{
Relative trajectories for pairs of spheres undergoing evolution in
shear flow (\protect\ref{external velocity}) in free space (top panel)
and in a channel of width $H/d=5$ (middle panel).  The spheres move in
the flow--gradient plane $x$--$z$, and the trajectories are shown in
the relative coordinates (\protect\ref{relative position}) for
different initial vertical offsets of particle centers $\zOffset$.  In
the wall-bounded system, the sphere with the larger initial value of
the coordinate $z$ starts at a distance $z/d=2.4$ from the lower wall
(as in figure \protect\ref{sample trajectories}). Bottom panel shows
the evolution of the dimensionless interparticle gap
(\protect\ref{gap}) for the trajectories displayed in the middle
panel.  The gap evolution for swapping trajectories is represented by
the heavy lines.
}
\label{collision map}
\end{figure}

\begin{figure}
  \centering
  \includegraphics[width=1\textwidth]{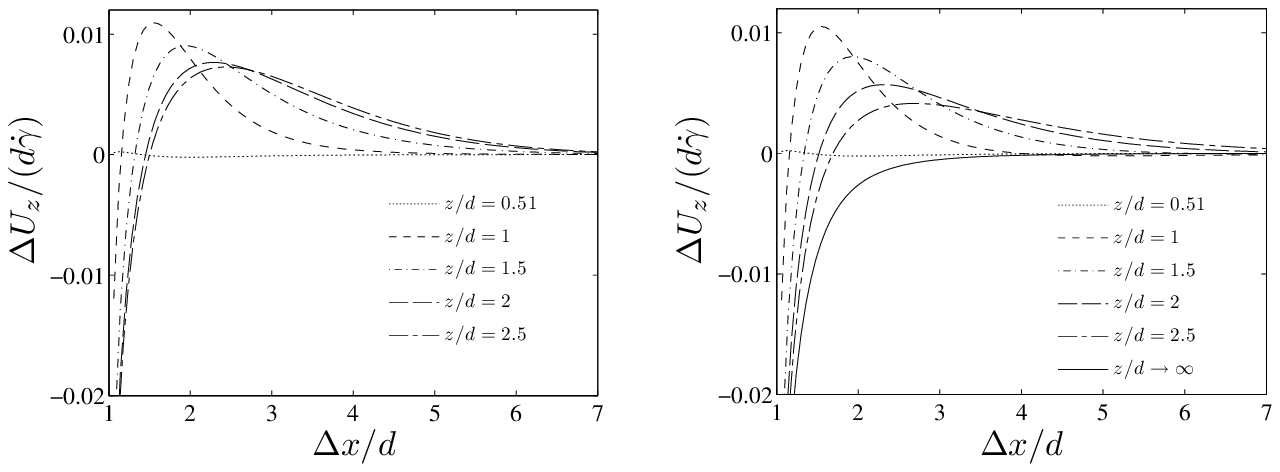}
\caption{
Vertical component of relative velocity (\protect\ref{relative
velocity}) for two spheres aligned along the flow direction $x$,
versus particle separation $\xOffset$.  Different lines correspond to
different positions $z$ of the particle pair with respect to the lower
wall. The left panel shows results for a two-wall system with wall
separation $H/d=5$ and the right panel for a one-wall system.
Positive values of $\relativeVelocity_z$ correspond to the
swapping-trajectory domain.
}
\label{signo vz}
\end{figure}

\section{Effect of walls on the dynamics of binary collisions}
\label{Effect of walls on binary collision dynamics}

\subsection{The morphology of open trajectories}
\label{The morphology of open trajectories}

An analysis
\citep*{Lin-Lee-Sather:1970,Batchelor-Green:1972a,Zinchenko:1984} of
the relative motion of a pair of spheres in unbounded shear flow
\refeq{external velocity} indicates that all open trajectories start
from $\xOffset=-\infty$ and extend to $\xOffset=+\infty$ (in a
reference frame fixed on the initially slower particle 1).  During the
approach of the particles their relative vertical offset $\zOffset$
increases, reaching a maximum when the particle pair crosses the
vorticity--gradient plane $\xOffset=0$.  However, after the particles
have separated, they return to their initial cross-streamline
positions $(\vorticityCoordinate,\gradientCoordinate)$, in accordance
with the flow-reversal symmetry of Stokes equations and the symmetry
of the system with respect to the reflection of the $\flowCoordinate$
coordinate.

For sufficiently large initial vertical offsets $\zOffset$ particle
trajectories in wall-bounded systems are qualitatively similar to those in
free space.  The only distinctive feature of the trajectory depicted
in the top panel of figure \ref{sample trajectories} is that $\zOffset$
decreases before reaching the maximum for the particle pair in the symmetry
plane $\xOffset=0$, while in free space $\zOffset$ would increase
monotonically.

However, for smaller initial values of $\zOffset$ we find an entirely
different behavior: for the new kind of open trajectories shown in the
bottom panel of figure \ref{sample trajectories} the offset $\zOffset$
changes sign before the relative position $\xOffset=0$ has been
reached.  Accordingly, the component $\Delta U_x$ of
the relative velocity
\begin{equation}
\label{relative velocity}
\relativeVelocityVector = \mathbf{U}_2-\mathbf{U}_1
\end{equation}
also changes sign.  The particles do not pass each other -- they turn
around and separate, maintaining $\xOffset<0$ for the whole
trajectory.  As a result, the spheres do not return to their initial
streamlines at long times but instead they swap their vertical
coordinates $z$.  Such particle encounters result, therefore, in
displacements of the suspended particles across streamlines of the
external flow.

The difference in topology of pair trajectories in unbounded space and
in a parallel-wall channel is clearly seen in figure \ref{collision
map}, where the relative particle motion is depicted in the reference
frame centered on one of the particles.  In free space (top panel) the
contact surface $|\Delta \br|=d$ is surrounded by an envelope of
closed orbits, and all open trajectories correspond to particles
passing each other.  In contrast, in the wall-bounded system (middle
panel) there also exists a region of swapping trajectories, delimited
by the critical trajectories crossing the plane $\Delta
z=0$ at the points where $\relativeVelocity_z=0$.

We emphasize that the swapping trajectories do not violate any
symmetries of the system.  Since the particles do not pass each other,
individual trajectories are not symmetric with respect to the
reflection $\flowCoordinate\to-\flowCoordinate$.  However, for each
trajectory in the halfspace $\xOffset<0$ there is a corresponding
reflected trajectory in the halfspace $\xOffset>0$. 

The position-swapping trajectories have several important
consequences.  First, they contribute to cross-streamline particle
migration in dilute suspensions bounded by planar or nearly planar
walls.  In particular, this migration mechanism explains the enhanced
hydrodynamic self-diffusivity observed by
\cite{Zarraga-Leighton:2002}, as discussed in section \ref{mixing
section}.

Another important consequence of the swapping mechanism is that it
prevents near-contact particle encounters.  The results depicted in
the middle panel of figure \ref{collision map} indicate that spheres
on the swapping trajectories approach each other less closely than
spheres on the usual trajectories (i.e.\ when the particles pass each
other).  A detailed view of the evolution of the dimensionless gap
between the particles,
\begin{equation}
\label{gap}
\gap=|\relativePosition|/d-1,
\end{equation}
is shown in the bottom panel of figure \ref{collision map}.  The results
(shown for each of the trajectories displayed in the middle panel) indicate
that the gap on non-swapping trajectories may decrease to about
$\gap\approx10^{-5}$, whereas on the swapping trajectories $\gap$ always
remains of order one.  Thus, the swapping mechanism may prevent particle
aggregation in the presence of short-range attractive forces.

\begin{figure}
  \centering
  \includegraphics[width=1\textwidth]{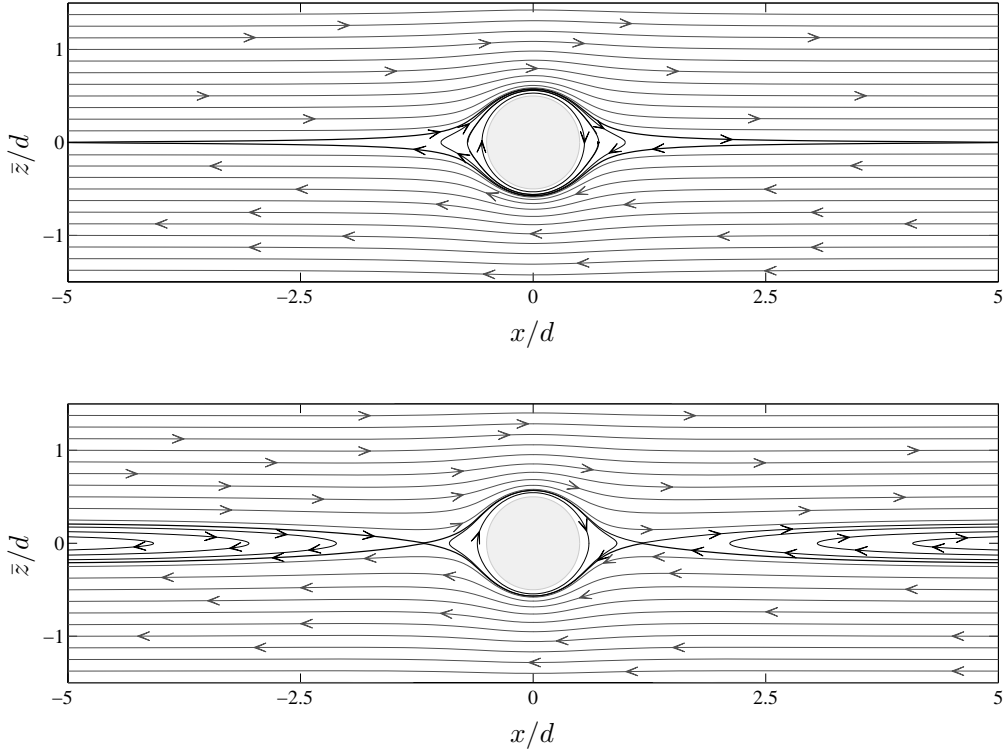}
\caption{
Fluid streamlines around a single sphere in external shear flow
(\protect\ref{external velocity}) in free space (top panel) and in a
channel of width $H/d=5$ (bottom panel).  Trajectories are depicted in
coordinate system centered on the sphere.  The particle in the channel
is in the midplane $z=H/2$.
}
\label{streamlines}
\end{figure}

\begin{figure}
  \centering
  \includegraphics[width=.55\textwidth]{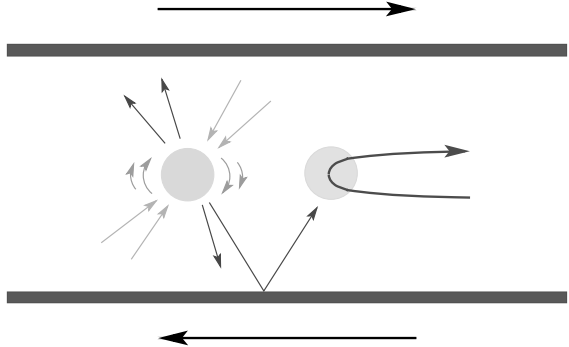}
\caption{
Explanation of the position-swapping mechanism: the wall reflection of
the stresslet flow produced by one of the spheres drives the other
sphere across the channel, causing reversal of the relative particle
motion.
}
\label{wall reflection sketch}
\end{figure}

\begin{figure}
  \centering
  \includegraphics[width=1\textwidth]{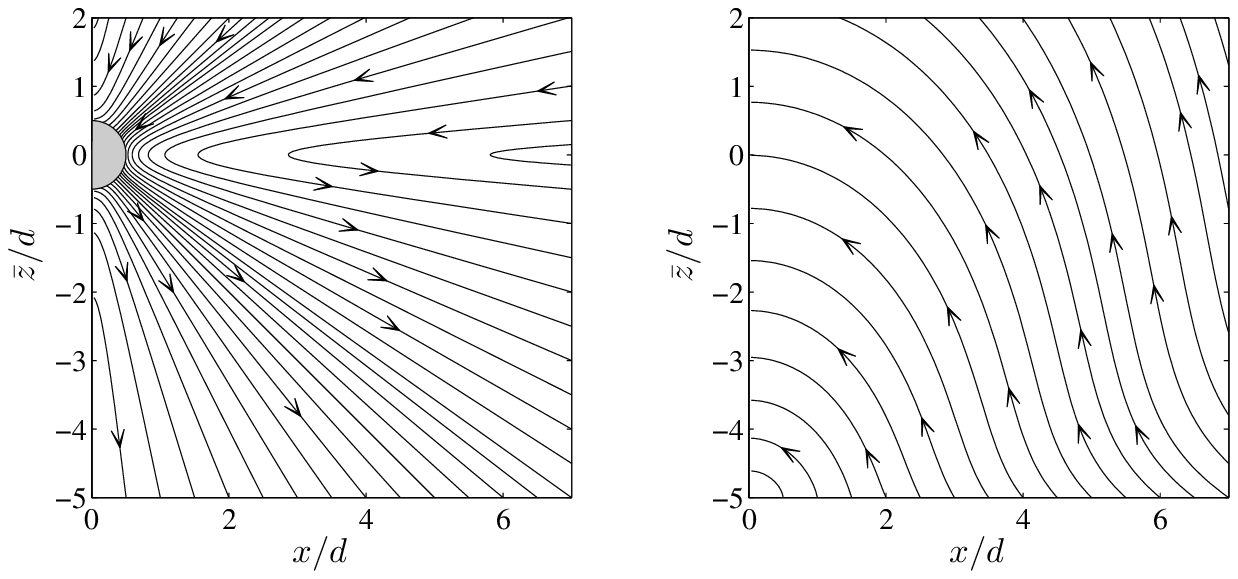}
\caption{
Scattered flow (\protect\ref{flow scattered by sphere in unbounded
shear flow}) produced by a sphere in shear flow (\protect\ref{external
velocity}) (left panel) and reflection $\scatteredFlow^*$ of this flow
from the wall located at $\zField/d=-5$ (right panel).  The direction
of the flow $\scatteredFlow^*$ is consistent with the swapping
mechanism illustrated in figure \protect\ref{wall reflection sketch}.
}
\label{plot of wall reflection}
\end{figure}

\subsection{Mechanism for particle swapping}
\label{Mechanism for particle swapping}

As shown in the middle panel of figure \ref{collision map}, the
swapping trajectories occur when the vertical component
$\relativeVelocity_z$ of the relative particle velocity
\refeq{relative velocity} is negative in a subdomain of particle
configuration with $\xOffset<0$ and positive in the corresponding
subdomain of the region $\xOffset>0$. Consider, for example, a pair of
spheres located in a flow--vorticity plane $z=\const$, and assume that
$\xOffset>0$, i.e., sphere 1 is to the left of sphere 2.  If
$\relativeVelocity_z>0$ in this configuration, sphere 2 moves from the
region below sphere 1 (where $\relativeVelocity_x<0$) to the region
above sphere 1 (where $\relativeVelocity_x>0$).  Thus, the particles
initially approach each other but then they separate without passing
each other -- they swap their vertical positions instead.

The dependence of the relative vertical velocity $\relativeVelocity_z$ on
particle separation $\xOffset$ is depicted in figure \ref{signo vz} for a
particle pair aligned in the flow direction.  The results indicate that in
wall-bounded systems the relative vertical particle velocity changes sign at a
critical particle separation $\xOffsetCrit$ that depends on the channel width
and the position of the particles with respect to the walls of the channel.
For $\xOffset>\xOffsetCrit$ the sign of $\relativeVelocity_z$ is consistent
with the topology of the swapping trajectories, and for
$\xOffset<\xOffsetCrit$ the sign of $\relativeVelocity_z$ corresponds to
closed orbits.  No such sign change, however, occurs in free space.

Although particle behavior on swapping trajectories seems at first
counterintuitive (the vector connecting the particle centers rotates
with an angular velocity opposite to the vorticity of the external
flow) it can be qualitatively explained by considering an analogy with
shear flow in a channel blocked by an obstacle.  If the channel is
completely blocked, the fluid dragged by a wall towards the obstacle
is deflected to the other side of the channel and returns along the
opposite wall.  If the channel is partly blocked (by one of the
spheres in our system) the streamlines that directly approach the
obstacle are deflected and exhibit the recirculation
behavior. Analogously, in a two-sphere system the swapping motion of
each particle is produced by the flow deflected by the other particle.

A more detailed, quantitative explanation of the swapping mechanism is
obtained by analyzing the effect of the walls on the flow pattern
$\bv_1$ around a single sphere in shear flow in a weakly confined
system.  To leading order in the particle--particle and
particle--wall separations, the fluid velocity $\bv_1$ and the
velocity of the second particle are identical.  Therefore, for weak
confinements the comparison of the particle and fluid velocities has a
quantitative meaning (assuming that the interparticle distance remains
sufficiently large), and for strongly confined systems, pair
trajectories are expected to qualitatively resemble the streamlines of
$\bv_1$.

The flow pattern around a single force- and torque-free sphere in an
unbounded and in a wall-bounded shear flow is illustrated in figure
\ref{streamlines}.  The streamlines are depicted in a the coordinate
system centered on the particle.  The top panel shows the familiar
velocity field for a particle in free space, and the bottom panel
represents the corresponding flow for a sphere in the middle of a
channel with wall separation $H=5d$.  As expected, in the wall-bounded
system there is a region of recirculating streamlines analogous to the
swapping trajectories shown in figure \ref{collision map}.

The hydrodynamic mechanism that leads to the recirculating streamlines
(and hence to the particle position-swapping) can be explained by
examining the wall reflection $\scatteredFlow^*$ of the perturbation
flow $\scatteredFlow$ produced by a sphere in external shear flow, as
schematically represented in figure \ref{wall reflection sketch}.
Only the reflection from the lower wall will be discussed because the
upper wall produces an analogous effect.  In our analysis we assume
that the wall is in the plane $z=0$ and the particle is at
$\br=(0,0,z_1)$, where $z_1/d\gg1$.

The perturbation flow $\scatteredFlow$ that results from
the scattering of the external flow \refeq{external velocity}
by the sphere can be expressed by the formula \citep{Kim-Karrila:1991}
\begin{equation}
\label{flow scattered by sphere in unbounded shear flow}
\scatteredFlow(\br)=-\smallfrac{1}{8}\shearRate d^3
   \left[\frac{5}{2}\frac{\xField\zField}{\rField^4}\rFieldUnitVector
   +\smallfrac{1}{8}d^2\bnabla\frac{\xField\zField}{\rField^5}\right],
\end{equation}
where $\rFieldVector=(x,y,\zField)$ (with $\zField=z-z_1$) is
the position of a field point $\br$ with respect to the particle
center, $\rField=|\rFieldVector|$, and
$\rFieldUnitVector=\rFieldVector/\rField$.  The flow $\scatteredFlow$
is a superposition of the $O(\rField^{-2})$ radial stresslet
contribution (the first term on the right-hand side of the above
equation) and the $O(\rField^{-4})$ potential contribution (the second
term).

The streamlines of the perturbation velocity field \refeq{flow
scattered by sphere in unbounded shear flow} are shown in the left
panel of figure \ref{plot of wall reflection} for a system with
particle-wall separation $z_1/d=5$.  In the plane $\zField=0$ (the
horizontal plane passing through the particle center), the radial
stresslet contribution vanishes, and the shorter-range potential-flow
contribution is vertical, with an orientation consistent with the
vorticity of the external flow.

Relation \refeq{flow scattered by sphere in unbounded shear flow} indicates
that at the wall surface (i.e.\ the plane $\zField=-z_1$) the perturbation
flow $\scatteredFlow$ points downwards for $x>0$ and upwards for $x<0$.  To
ensure that there is no fluid flux through the wall, the flow reflected from
the wall, $\scatteredFlow^*$, must point in the opposite direction.
Therefore, assuming that the vertical component of $\scatteredFlow^*$ is a
monotonic function of $z$ (at least up to the particle position), we find that
the reflected flow $\scatteredFlow^*$ has the direction needed to produce the
recirculating streamlines (cf. figure \ref{wall reflection sketch}).

It can be shown that $\scatteredFlow^*$ decays as $O(\rFieldR^{-2})$ with the
distance $\rFieldR$ from the image singularity at $\br=(0,0,-z_1)$ (the image
flow involves the reflection of the stresslet).  It follows that for
$\zField=0$ the flow $\scatteredFlow^*$ dominates the oppositely directed
$O(\rField^{-4})$ potential-flow term of $\scatteredFlow$, except in the
region near the particle.  This behavior is consistent with the pattern of the
streamlines depicted in the bottom panel of figure \ref{streamlines} and with
the topology of pair trajectories represented in the middle panel of figure
\ref{collision map}.

The above essential features of the reflected flow can be explicitly
seen for a simple case of free interface.  To the leading order in
$z_1/d$, the flow reflected from such an interface is given by the
expression
\begin{equation}
\label{reflection of stresslet}
\scatteredFlow^*(\br)=\shearRate d^3
   \frac{5}{16}\frac{\xFieldR\zFieldR}{\rFieldR^4}\rFieldUnitVectorR,
\end{equation}
where $\zFieldR=z+z_1$ and $\rFieldVectorR=(x,y,\zFieldR)$ denote the
$z$ coordinate and position vector in the reference frame centered at
the image singularity.  Note that for positions close to
the particle, the magnitude of the flow \refeq{reflection of
stresslet} is $O(d^2x/\rFieldR^3)$, because both $d\ll\rFieldR$ and
$x\ll \rFieldR$.

For rigid walls, the crucial features of the reflected flow (i.e., the
direction consistent with the swapping particle motion, the overall
$(d/\rFieldR)^2$ decay, and the small additional factor
$O(x/\rFieldR)$ for $\scatteredFlowComponent{z}^*$ near the plane
$x=0$) can be derived using \cite{Lorentz:1907} reflection formula, as
discussed in Appendix \ref{Flow reflected from rigid wall}.  These
features can also be seen from the streamlines of $\scatteredFlow^*$
depicted in the right panel of figure \ref{plot of wall reflection}.

In the above analysis of the particle-swapping mechanism we have
focused on a single reflection $\scatteredFlow^*$ of the perturbation
flow from a wall.  However, complete multi-body hydrodynamic particle
interactions in wall presence are accurately accounted for in our
quantitative calculations.  While our paper is focused on particle
motion in the creeping flow regime we would also like to note that
fluid reversal zones and position-swapping pair trajectories similar
to those seen in figures \ref{collision map} and \ref{streamlines}
were observed at finite Reynolds numbers for unconfined configurations
\citep{%
Mikulencak-Morris:2004,%
Subramanian-Koch:2006,%
Kulkarni-Morris:2007%
}.

\begin{figure}
  \centering
  \includegraphics[width=.5\textwidth]{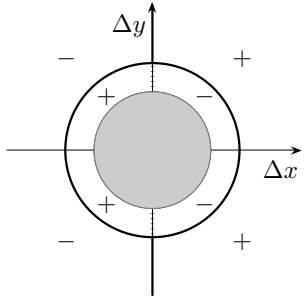}
\caption{
Location of zeros of the relative vertical particle velocity
$\relativeVelocity_z$ and sign of $\relativeVelocity_z$ for two
spheres in the midplane $z=H/2$ of a channel with wall separation
$H/d=5$.  Shaded region (of diameter $d$) corresponds to overlapping
spheres.  Solid lines indicate location of zeros that delimit the
region of swapping trajectories, and dotted lines represent zeros in
the periodic-orbit domain.
}
\label{signo vz offset A}
\end{figure}

\begin{figure}
  \centering
  \includegraphics[width=1\textwidth]{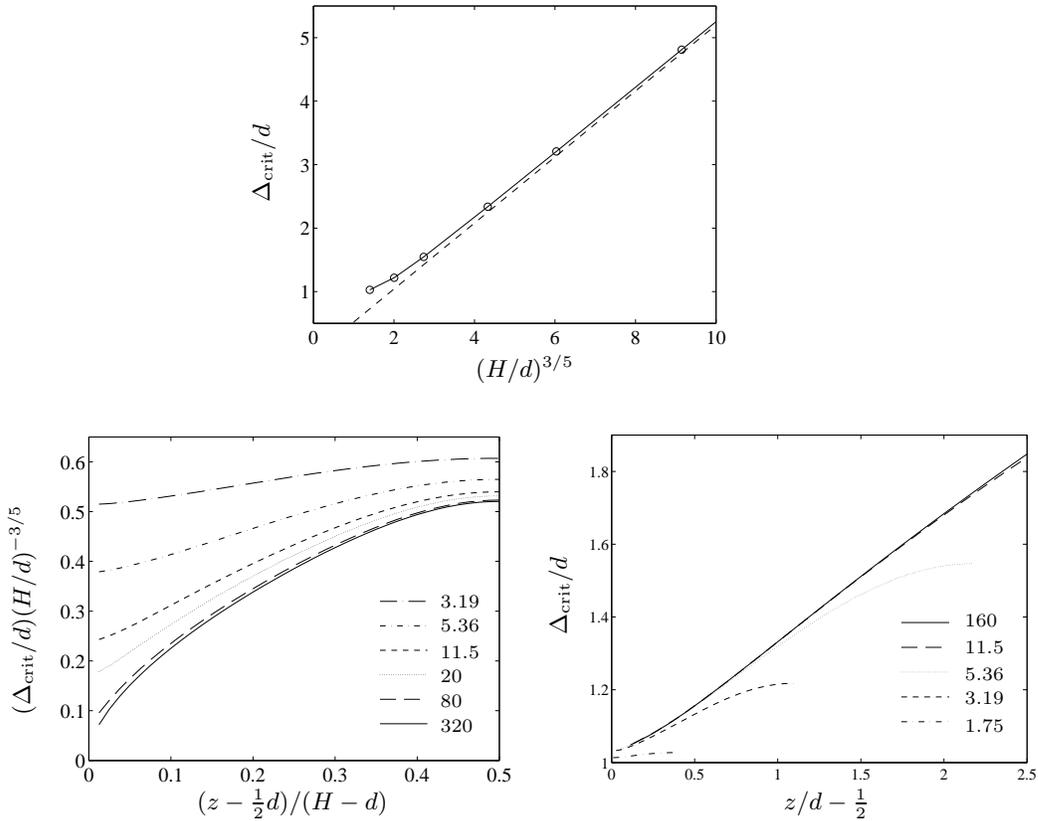}
\caption{
Radius $\rhoOffsetCrit$ of the circle of zeros of
$\relativeVelocity_z$.  Top panel shows the normalized radius versus
channel width $H$ for a particle pair in the midplane $z=H/2$.  Solid
line represents numerical results and dashed line the asymptotic
scaling (\protect\ref{scaling of Delta crit}).  Bottom panels show
$\rhoOffsetCrit$ versus position $z$ of the particle pair for several
channel widths $H/d$ (as labeled).  The results in the left panel are
scaled to emphasize asymptotic behavior (\protect\ref{scaling of Delta
crit}), and the results in the right panel are shown unscaled to
emphasize the near-wall behavior.
}
\label{signo vz offset B}
\end{figure}

\begin{figure}
  \centering
  \includegraphics[width=1\textwidth]{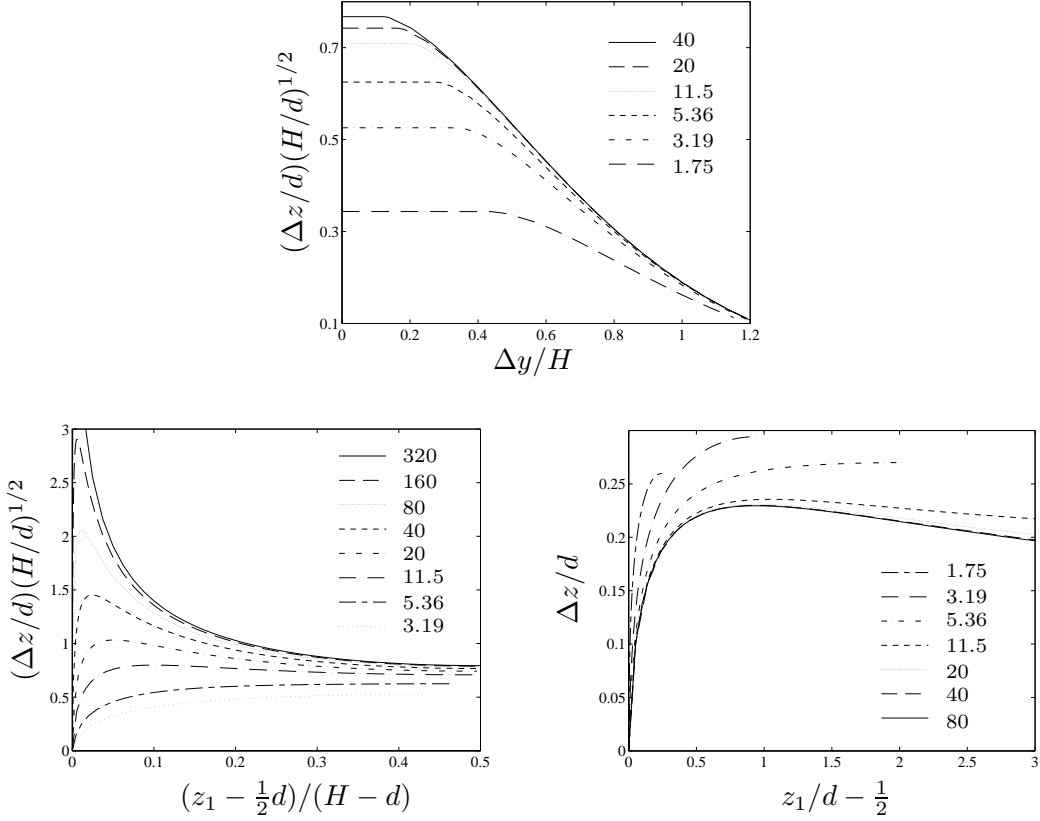}

\caption{
Upstream domains (in transverse coordinates) corresponding to swapping
trajectories, for different values of the channel width (as labeled).
The top panel depicts these domains in coordinates
$(\yOffset,\zOffset)$ for particles moving symmetrically with respect
to the midplane $z=H/2$.  The bottom panels show the relative vertical
offset $\zOffset$ versus the upstream vertical position $z_1$ of the
particle closer to the wall, for particles moving in the
flow--gradient plane $\yOffset=0$.  The results in the top and left
bottom panels are rescaled to emphasize the asymptotic behavior
(\protect\ref{z Offset for large H}), and the results in the bottom
right panel are shown unscaled to emphasize the near-wall behavior.  The
curves represent the initial coordinates of the limiting trajectories,
and the regions below the curves correspond to the swapping trajectory
domains.
}
\label{boundary lines A}
\end{figure}

\begin{figure}
  \centering
  \includegraphics[width=.5\textwidth]{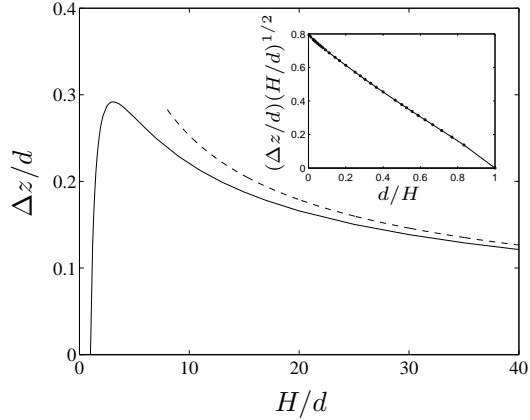}
\caption{
Vertical extent $\zOffset$ of the upstream domains of swapping
trajectories for particles moving symmetrically with respect to the
midplane $z=H/2$ (as in the top panel of figure \protect\ref{boundary
lines A}), plotted versus the channel width $H$.  Solid line
represents numerical results and dashed line the
asymptotic behavior (\protect\ref{z Offset for large H}).  The results
in the inset are rescaled accordingly.
}
\label{boundary lines B}
\end{figure}

\begin{figure}
  \centering
  \includegraphics[width=1\textwidth]{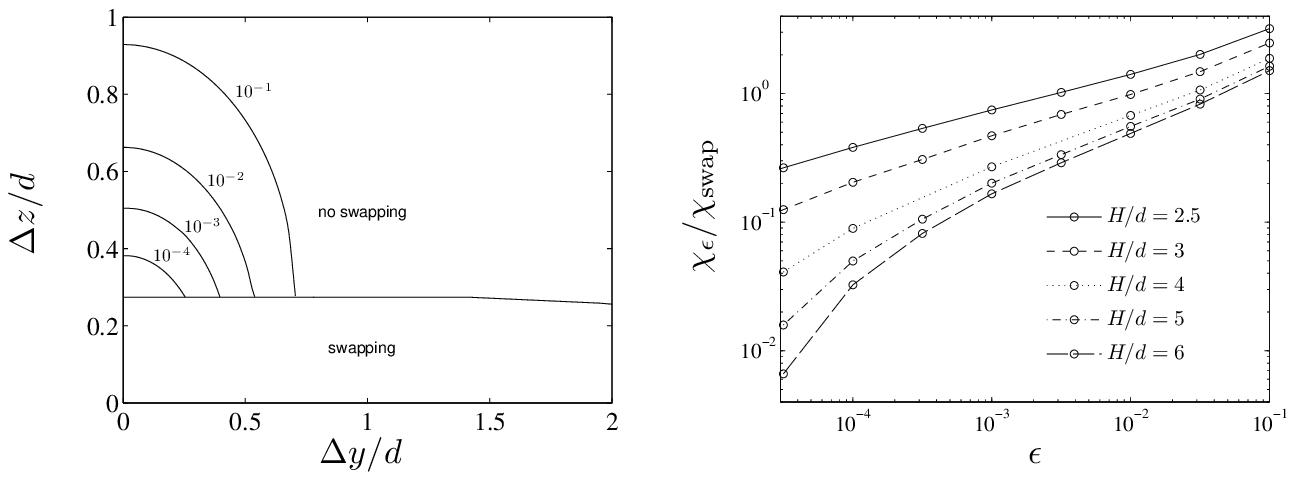}
\caption{
Left panel represents domains in the upstream transverse coordinates
$(\yOffset,\zOffset)$, corresponding to several values of the minimal gap
$\gap$ (as labeled) for a two-wall system with wall separation $H/d=5$.  Right
panel shows the ratio between the frequency $\chiGap$ of near contact particle
encounters and the frequency $\chiSwap$ of swapping-trajectory encounters in a
dilute suspension under shear, versus the cutoff value of the minimal gap
$\gap$ for several wall separations $H/d$.  All results are for trajectories
symmetric with respect to the midplane $z=H/2$. 
}
\label{fixed gap cross section}
\end{figure}

\subsection{The domain of swapping trajectories}
\label{Characterization the domain of swapping trajectories}

A pair of particles on a swapping trajectory crosses the horizontal plane
$\zOffset=0$ in the direction corresponding to the counter-rotation of the
vector connecting the particle centers (cf. section \ref{Mechanism for
particle swapping}).  Since the counter-rotation ceases at the points where
$\relativeVelocity_z$ changes sign, the region of swapping trajectories is
delimited by a set of critical trajectories for which
$\relativeVelocity_z\to0$ when the particle pair approaches the plane
$\zOffset=0$.  (We note that for $\zOffset=0$ the condition
$\relativeVelocity_z=0$ implies that $\relativeVelocityVector=0$ because the
horizontal components of the relative velocity \refeq{relative velocity}
vanish because the symmetry of the problem.)

For a two-wall system a typical graph showing the sign of the vertical
relative velocity $\relativeVelocity_z$ and the location of points
where $\relativeVelocity_z=0$ in the flow--vorticity plane
$\zOffset=0$ is presented in figure \ref{signo vz offset A}.  The set
of points where $\relativeVelocity_z$ vanishes consists of two
subsets.  One is the line $\xOffset=0$ in the region with no particle
overlap, and the other is the circle $\rhoOffset=\rhoOffsetCrit$
(where $\rhoOffsetVector=\xOffset\hat\be_x+\yOffset\hat\be_y$ denotes
the horizontal relative position vector,
$\rhoOffset=|\rhoOffsetVector|$, and $\rhoOffsetCrit$ is the circle
radius).  The critical interparticle distance $\rhoOffsetCrit$ does
not depend on the orientation of the particle pair in the plane
$\zOffset=0$ due to symmetry (cf.\ explanation in Appendix \ref{New
appendix}).

By analyzing the direction of sphere motion, one can show that the
area inside the circle of zeros $\rhoOffset=\rhoOffsetCrit$
corresponds to closed trajectories passing through the plane
$\zOffset=0$.  Therefore, only the part of the line $\xOffset=0$
outside this circle is associated with the critical trajectories that
delimit the swapping region.  We note that closed orbits also exist in
unbounded shear flow \citep{Batchelor-Green:1972a}.  In this case the
whole surface $\zOffset=0$ corresponds to particle pairs on closed
trajectories (as in the top panel of figure~\ref{collision map}).

For weakly confined systems the location of zeros of
$\relativeVelocity_z$ results from the balance between the
$\sim(\rhoOffset/d)^{-4}$ potential-flow contribution in equation
\refeq{flow scattered by sphere in unbounded shear flow} and the wall
reflection of the stresslet.  As already indicated in section
\ref{Mechanism for particle swapping} (cf.\ also Appendix \ref{Flow
reflected from rigid wall}), the leading-order behavior of the
vertical component of the reflected flow in the region near the
particle is
\begin{equation}
\label{leading-order behavior of image flow}
\scatteredFlowComponentW{z}\sim\left(\frac{d}{H}\right)^2\frac{\rhoOffset}{H}.
\end{equation}
Thus, for a fixed position $z/H$ of the particle pair in the channel
we find that
\begin{equation}
\label{scaling of Delta crit}
\rhoOffsetCrit/d\sim(H/d)^{3/5},\qquad H/d\gg1.
\end{equation}

The dependence of the radius $\rhoOffsetCrit$ on the channel width for
a particle pair in the midplane $z/H=\halff$ is plotted along with the
asymptotic behavior \refeq{scaling of Delta crit} in the top panel of
figure \ref{signo vz offset B}.  The two bottom panels of figure
\ref{signo vz offset B} show the dependence of $\rhoOffsetCrit$ on the
position of the particle pair with respect to the channel walls for
different $H$.  The results represented in the left panel correspond
to larger values of $H/d$, and they are rescaled to emphasize the
asymptotic behavior \refeq{scaling of Delta crit}.  In the right panel
the results are shown unscaled for moderate values of $H/d$.  Note
that for sufficiently small values of $z/H$ the unscaled curves
corresponding to different channel widths coincide, which indicates
that the particle mobility is dominated by the single-wall
contribution except for particles close to the center of the channel.

An alternative way of characterizing the domain of swapping
trajectories is to show  the corresponding upstream
region of the transverse relative coordinates $(\yOffset,\zOffset)$,
i.e., the region through which the swapping trajectories pass in the
limit of infinite streamwise particle separations
$\xOffset\to-\infty$.  For trajectories that cross a given horizontal
plane $\zOffset=0$ during the particle motion, the border of the
upstream region can be determined by integrating the trajectories
backwards, starting near the loci of zeros of $\relativeVelocity_z$ in
the plane $\zOffset=0$.

In figure \ref{boundary lines A} the upstream boundaries of the
swapping-trajectory regions are shown for several values of the wall
separation $H/d$.  The top panel depicts these boundaries for the trajectories
that are symmetric with respect to the midplane $z=H/2$ (strictly speaking,
symmetric with respect to the axis defined by the intersection of the planes
$y=(y_1+y_2)/2$ and $z=H/2$).  The bottom panels illustrate the dependence of
the vertical extent of the upstream regions $\zOffset$ on the initial position
$z_1$ of the lower particle for particle pairs moving in the flow--gradient
plane $\yOffset=0$.  In all the panels the swapping-trajectory domains
correspond to the areas below the curves.

By the scaling argument outlined in Appendix~\ref{Appendix}, one can
show that the vertical extent of the upstream swapping-trajectory
region scales as
\begin{equation}
\label{z Offset for large H}
\zOffset/d\sim(H/d)^{-1/2},\qquad H/d\gg1,
\end{equation}
which is confirmed by the results shown in figure \ref{boundary lines
B}.  The vertical axes of the plots shown in the top and left bottom
panels of figure \ref{boundary lines A} are scaled accordingly, to
collapse the results for large values of $H/d$ onto the asymptotic
master curves.  The results in the right bottom panel are shown
unscaled to emphasize the near-wall behavior of the swapping
trajectory domain.

An interesting feature of the swapping-trajectory regions shown in the
top panel of figure \ref{boundary lines A} is that these domains are
delimited by straight horizontal lines for $\yOffset\le
\yOffset_\crit\approx\rhoOffsetCrit$.  The straight-line sections
correspond to the trajectories that pass through the 
circle of zeros $\rhoOffset=\rhoOffsetCrit$ of the vertical relative
velocity $\relativeVelocity_z$ when the particle pair crosses the
horizontal plane $\zOffset=0$.  The remaining portions of the upstream
borders of the swapping-trajectory regions correspond to the zeros of
$\relativeVelocity_z$ located at the axis $\xOffset=0$ in the plane
$\zOffset=0$ (cf.\ the geometry of zeros of $\relativeVelocity_z$
depicted in figure~\ref{signo vz offset A}).  The upstream coordinate
$\zOffset$ in the region $\yOffset\le \yOffset_\crit$ is independent
of the corresponding position along the circle of zeros of
$\relativeVelocity_z$ for the same reason why $\rhoOffsetCrit$ does
not depend on the orientation angle $\phi$ (cf.\ Appendix \ref{New
appendix}).

The results in the top panel of figure \ref{boundary lines A} indicate that
the lateral extent $\yOffset/H$ of the upstream swapping-trajectory region
(normalized by the wall separation $H$) remains approximately constant when
the wall separation $H$ increases.  However, the horizontal coordinate
$\yOffset_\crit/H$ of the end point of the straight-line section decreases, as
required by the scaling \refeq{scaling of Delta crit}.

We conclude this section by discussing the distance of minimal interparticle
approach for pairs of spheres on swapping and non-swapping trajectories.  As
we have already indicated, the interparticle gap \refeq{gap} on trajectories
of the swapping kind typically remains relatively large.  For spheres that
pass through a given horizontal plane $\zOffset=0$, the minimal interparticle
separation on swapping trajectories coincides with the radius $\rhoOffsetCrit$
of the circle of zeros of the relative transverse velocity
$\relativeVelocity_z$.  According to the results shown in the top panel of
figure \ref{signo vz offset B}, $\rhoOffsetCrit$ is typically well above the
particle diameter $d$ (except for very small particle--wall separations).
Thus, the minimal interparticle gap $\gapMinSwap=\rhoOffsetCrit/d-1$ is
usually $O(1)$.  In contrast, the minimal interparticle gap on non-swapping
open trajectories may be smaller by several orders of magnitude.  The sharp
transition from the large gaps on the swapping trajectories to much smaller
values of $\gap$ on the nearby non-swapping ones is illustrated in the bottom
panel of figure \ref{collision map}.

The domains of non-swapping trajectories corresponding to specific
values of the minimal gap are depicted in the left panel of figure
\ref{fixed gap cross section} for a system with a wall separation
$H/d=5$ and particles moving symmetrically with respect to the
midplane $z=H/2$.   The results indicate that the smallest values of the
minimal gap are attained for near-critical non-swapping trajectories
in the flow--gradient plane $\yOffset=0$.  According to our additional
calculations the minimal gap decreases for stronger confinements,
although it typically remains comparable to the distance of minimal
approach $\gap_{\textrm{min}}=4.2\times10^{-5}$ for two spheres on
open trajectories in the unbounded shear flow \citep{Arp-Mason:1977a}.

In a system of spheres interacting via a short-range repulsive potential, the
near-contact particle encounters may result in cross-streamline particle
displacements.  However, for moderate-width channels, the upstream region
corresponding to very small gaps is significantly smaller than the upstream
region of swapping trajectories.  

Quantitatively, the rate $\chi$ of particle encounters of a given type
can be evaluated by integrating the upstream relative particle
velocity $\relativeVelocity_\infty$ over the appropriate
upstream-coordinate region.  The ratio
$\chi_{\gap}/\chi_{\textrm{swap}}$ between the rates of near-contact
and swapping collisions for trajectories symmetric with respect to the
midplane $z=H/2$ is plotted in the right panel of figure \ref{fixed
gap cross section} versus the cutoff value of the minimal gap for
channels of different width.  The results indicate that
$\chi_{\gap}/\chi_{\textrm{swap}}\ll1$ for sufficiently small values
of the gap, especially at moderate confinements.

\begin{figure}
  \centering
  \includegraphics[width=1\textwidth]{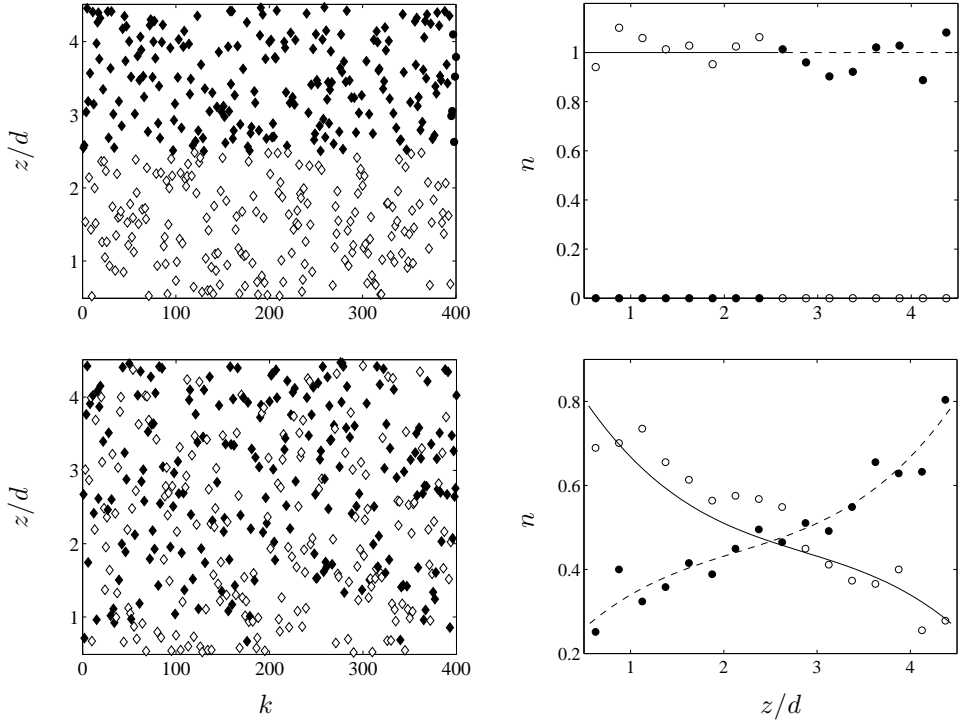}
\caption{
Particle migration in a dilute binary suspension of equal-size spheres
undergoing shear flow in a parallel-wall channel with wall separation
$H/d=5$.  Top panels show the initial state of the system, and bottom
panels represent the state after 128 swapping binary encounters per
particle.  The left panels depict vertical positions $z/d$ for
randomly selected $200$ particles of each species (versus the particle
index $k$), and right panels show the normalized particle density $n$
across the channel.  The curves in right-bottom panel represent the
$3^{\textrm{rd}}$ order polynomial fit to the histogram of the
particle distribution (taking the system symmetry into account)
}
\label{mixing}
\end{figure}

\begin{figure}
  \centering
  \includegraphics[width=1\textwidth]{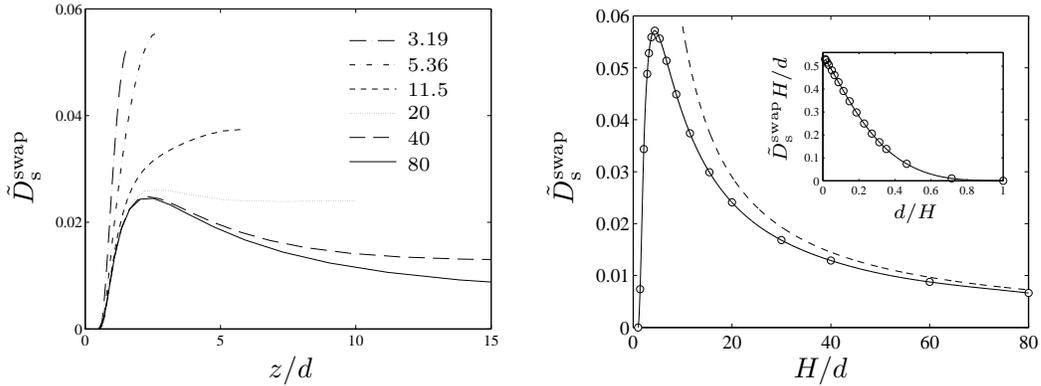}
\caption{
Swapping-trajectory contribution (\protect\ref{self-diffusivity swap
-- non-dimensional integral}) to the hydrodynamic self-diffusivity
coefficient.  Left panel shows $\selfDiffusivitySwapND$ versus
distance $z/d$ from the lower wall for different values of the channel
width $H/d$ (as labeled).  Right panel shows $\selfDiffusivitySwapND$
at the center of the channel versus $H/d$.  Dashed line
represents the asymptotic behavior (\protect\ref{scaling of
diffusivity with gap}), and the inset shows the results in the
corresponding rescaled variables.}
\label{plots of self-diffusivity}
\end{figure}

\section{Particle migration due to position-swapping binary encounters }
\label{mixing section}

As shown above, the position-swapping particle encounters result in
cross-streamline particle migration in suspensions confined by planar
(or nearly planar) walls.  Although this migration mechanism may be
dominant in dilute suspensions under mode\-ra\-te-confinement
conditions, it has never been described.  In this section we present
our quantitative predictions for migration of spherical particles in
parallel wall channels.

\subsection{Population-balance simulations}
\label{Population-balance simulations} 

In order to illustrate the role of particle swapping in suspension flows, we
have performed a population-balance simulation of a mixture of two species of
equal-size spheres in a parallel-wall channel of width $H/d=5$.  Initially,
the particles of species one are located in the upper portion of the channel,
$z>H/2$, and the particles of species two in the lower portion $z<H/2$.  In
each part of the channel the particles are distributed randomly.  Since the
system is translationally invariant in the transverse directions $x$ and $y$,
the particle distribution is characterized only by the vertical coordinate
$z$.

To mimic the evolution of a dilute suspension, the particle population
is updated using a Boltzmann--Monte Carlo technique.  Accordingly, a
pair of particles is chosen randomly from the ensemble representing
the system, with a probability proportional to the upstream relative
velocity $\relativeVelocity_\infty$.  Then, a two-particle evolution
run is performed with a large initial offset in the streamwise
direction $|\xOffset|\gg d$ and the transverse offset $\yOffset$
chosen randomly from the uniform probability distribution truncated at
$\yOffset_{\textrm{max}}\approx 8d$.  If the pair evolution results in
position swapping, the particle ensemble is updated accordingly.

Sample results of our population-balance simulations are presented in
figure \ref{mixing}.  The calculations were performed using a set of
600 particles, and the average over seven simulation runs was taken to
reduce statistical fluctuations.  The top panels show the initial
random particle distribution, and the bottom panels the distribution
after 128 swapping collisions per particle.  In the left panels the
distribution is depicted by plotting the vertical positions of 400
randomly chosen particles versus particle index.  The right panels
represent the corresponding particle density (normalized to unity) as
a function of the position $z$ across the channel.  The results
indicate that the two particle species slowly mix together due to the
particle-swapping phenomenon.  Without particle swapping, all spheres
would preserve their initial vertical positions (unless
non-hydrodynamic interparticle forces are present).

\subsection{Self-diffusion coefficient}
\label{Self-diffusion coefficient}

Assuming that the wall separation $H$ is much larger than the swapping range
$\zOffset$, a sequence of uncorrelated particle displacements due to binary
encounters in suspension flow can be approximately described as a diffusion
process. In this section we focus on the transverse diffusivity of a tagged
particle in a dilute suspension of mechanically identical spheres.

As shown by \cite{da_Cunha-Hinch:1996} in their paper on the
shear-induced self-diffusivity in a dilute suspension of rough spheres
\cite*[also see][]{Zarraga-Leighton:2001}, the transverse
(cross-streamline) self-diffusion coefficient $\selfDiffusivity$, can
be evaluated from the relation
\begin{equation}
\label{self-diffusivity general formula}
\selfDiffusivity=\halff n\int\int(\Delta Z)^2\relativeVelocity_\infty 
      \diff\yOffset\diff\zOffset,
\end{equation}
where the integration is over the upstream region of transverse
relative coordinates corresponding to particle encounters of a given
type (e.g., swapping or direct particle collisions), $n$ is the
particle number density and $\Delta Z$ is the cross-streamline
displacement of the particle during a binary-encounter event.

Since initially the particles are well separated in the streamline
direction, the relative upstream velocity
$\relativeVelocity_\infty=U_\infty(z_2)-U_\infty(z_1)$ is the
difference of the velocities $U_\infty(z)$ of individual,
non-interacting spheres at the positions $z_1$ and $z_2$ with respect
to the channel walls.   For particles far from the walls we
simply have $\relativeVelocity_\infty=\shearRate\zOffset$.  More
generally, assuming that the initial particle offset $\zOffset$ is
sufficiently small, the relative particle velocity
$\relativeVelocity_\infty$ can be expressed as
\begin{equation}
\label{upstream relative velocity}
\relativeVelocity_\infty=\shearRate\alpha\zOffset,
\end{equation}
where
\begin{equation}
\label{derivative of one-particle velocity}
\alpha(z)=\shearRate^{-1}\frac{\diff U_\infty}{\diff z}
\end{equation}
is the dimensionless derivative of the velocity of an individual
sphere with respect to its transverse position in the channel.
According to our results
\citep{Zurita_Gotor-Blawzdziewicz-Wajnryb:2007a}, the velocity of a
single sphere undergoing shear flow in a parallel-wall channel differs
very little from the local fluid velocity \refeq{external velocity},
except when the particle is in the immediate wall proximity.
Therefore, the approximation $\alpha=1$ is often sufficient.

Particles on swapping trajectories exchange their vertical positions, which
implies that the particle displacement $\Delta Z$ equals the upstream offset
$\zOffset$ of the particles in the interacting pair.  Using the above result
and relation \refeq{upstream relative velocity}, the integral with respect to
$\zOffset$ in equation \refeq{self-diffusivity general formula} can be
explicitly performed over the swapping trajectory region.  The effective
self-diffusivity can thus be determined from the resulting one-dimensional
integral
\begin{equation}
\label{self-diffusivity swap}
\selfDiffusivitySwap=\smallfrac{1}{4} n\shearRate\alpha\int_{-\infty}^\infty 
    \zOffsetMax^4 \diff\yOffset,
\end{equation}
where $\zOffsetMax=\zOffsetMax(\yOffset)$ is the upstream vertical
particle offset on the trajectory that delimits the region of swapping
trajectories (cf.\ the contour plots in the top panel of figure
\ref{boundary lines A}).  Rephrased in terms of the volume fraction
$\phi=\frac{1}{6}\pi d^3n$ and dimensionless particle offsets
$\yOffsetND=\yOffset/d$ and $\zOffsetND=\zOffset/d$, equation
\refeq{self-diffusivity swap} yields
\begin{equation}
\label{self-diffusivity swap -- non-dimensional definition}
\selfDiffusivitySwap=\quarter\shearRate\phi d^2\selfDiffusivitySwapND,
\end{equation}
where
\begin{equation}
\label{self-diffusivity swap -- non-dimensional integral}
\selfDiffusivitySwapND=6\pi^{-1}\alpha
    \int_{-\infty}^\infty \zOffsetMaxND^4 \diff\yOffsetND.
\end{equation}

The dimensionless self-diffusion coefficient $\selfDiffusivitySwapND$
depends on the position $z$ of the particle pair across the channel.
Since the problem is non-local, we assign $z$ to be the position of
the particle pair when it crosses the horizontal plane $\zOffset=0$.

Our numerical results for the swapping contribution
\refeq{self-diffusivity swap -- non-dimensional integral} to the
effective self-diffusion coefficient are presented in figure
\ref{plots of self-diffusivity}.  The left panel shows
$\selfDiffusivitySwapND$ versus $z$ for several values of channel
widths, and the right panel shows the self-diffusivity at the channel
center as a function of $H$.  The results of our calculations indicate
that for $H/d\lesssim 15$ the self-diffusion coefficient
$\selfDiffusivitySwapND$ achieves its maximal value at the channel
center, where the swapping effect is the strongest due to the
superposition of the flow reflected from two walls.  In contrast, for
$H/d\gtrsim15$ the maximum occurs at the distance $z/d\approx 2$ from
each wall.  The results also show that particle migration is
suppressed near the walls, which is consistent with the observation
that the vertical extent of the particle-swapping domain is small near
the walls (cf.\ the bottom panels of figure \ref{boundary lines A}).
The effect of the decreased self-diffusivity in the near-wall regions
is reflected in the shapes of the density profiles depicted in figure
\ref{mixing}.

The behavior of $\selfDiffusivitySwapND$ for $H/d\gg1$ can be
determined by combining relations \refeq{z Offset for large H} and
\refeq{self-diffusivity swap -- non-dimensional integral} and recalling
that the lateral extent $\yOffset$ of the swapping region scales with
$H$ for the large channel width (as depicted in the top panel of
figure \ref{boundary lines A}).  Accordingly, we obtain the relation
\begin{equation}
\label{scaling of diffusivity with gap}
\selfDiffusivitySwapND=q\dot\gamma\phi d^3/H,
\end{equation}
where $q$ depends on the position of the particle pair across the
channel $z/H$.  For particles in the midplane $z/H=\halff$ we find
$q\approx0.55$, according to the results shown in the inset of the
right panel of figure \ref{plots of self-diffusivity}.

\subsection{Comparison with experiment  by Zarraga and Leighton}
\label{Comparison with experiment}

While the mechanism for particle migration due to the swapping trajectories
has not been proposed so far, the effect of the swapping motions was observed
when \cite{Zarraga-Leighton:2002} reported unusually large values of the
hydrodynamic self-diffusivity for a dilute suspension of spheres undergoing
shear flow in a Couette device.  Their measurements gave the result
$\selfDiffusivity=3.6\times10^{-2}\,\shearRate\phi a^2$ (where $a=\halff d$ is
the particle radius) for the low-density self-diffusion coefficient, whereas
the estimate of the contribution due to particle roughness was at least four
times smaller: it ranged from
$\selfDiffusivityRough=3.0\times10^{-3}\,\shearRate\phi a^2$ to
$8.4\times10^{-3}\,\shearRate\phi a^2$, depending on the assumed roughness
amplitude.  This discrepancy has been unaccounted for, and the result of
\cite{Zarraga-Leighton:2002} experiment has been puzzling.

By analyzing the effect of the channel walls on particle motion we
provide a simple explanation for the anomalous value of the
self-diffusion coefficient.  The \cite{Zarraga-Leighton:2002}
measurements were performed in a channel with wall separation
$H/d=20$.  For this geometry we find that the swapping contribution to
the self-diffusion coefficient in the center of a channel is
$\selfDiffusivitySwap=2.4\times10^{-2}\,\shearRate\phi a^2$.
Moreover, in the whole central region the diffusivity
$\selfDiffusivitySwap$ only weakly depends on the position $z$,
according to the results shown in figure \ref{plots of
self-diffusivity}.  Assuming the upper limit of the roughness
parameter $\roughnessParameter=1.8\times10^{-2}$ quoted by
\cite{Zarraga-Leighton:2002}, we find
$\selfDiffusivityRough=6.7\times10^{-3}\,\shearRate\phi a^2$ (note
that $\selfDiffusivityRough$ is slightly smaller than the
corresponding value in infinite space because the swapping mechanism
reduces the rate of direct particle collisions).  Combining the
swapping and roughness contributions, we obtain
$\selfDiffusivity=3.1\times10^{-3}\,\shearRate\phi a^2$ for the total
self-diffusion coefficient.  Our result agrees with the experimental
value with accuracy of 15\,\%.

In this paper we focus on systems of equal-size spheres but wall-induced
cross-streamline particle migration also occurs in bidisperse or polydisperse
suspensions.  We note that in such suspensions, migration resulting from
particle roughness is significantly smaller than in monodisperse ones
\citep{Zarraga-Leighton:2001,Zarraga-Leighton:2002}.  Thus in polydisperse
systems particle swapping constitutes the dominant particle migration
mechanism even if there is significant particle roughness.

Moreover, cross-streamline displacements of two different-size
particles undergoing a binary swapping encounter differ because there
is no fore--aft symmetry.  It follows that the total suspension
density is affected by the swapping trajectories, in addition to the
migration of individual particle species.  In contrast, in
monodisperse suspensions the total density cannot be altered by
position-swapping binary encounters.

\section{Conclusions}
\label{Conclusions}

The key finding of our study is that confining walls can qualitatively
change the topology of binary encounters of spherical particles in
suspension flows.  Specifically, we have identified a new class of
binary trajectories that result in cross-streamline particle migration
in a wall-bounded shear flow.  Equal-size spheres on such trajectories
do not pass each other (as in an unbounded system) but, instead, they
exchange their transverse positions.  While our explicit calculations
are limited to equal-size spherical particles in shear flow bounded by
parallel planar walls, we expect that a similar effect also exists for
pressure-driven flows, particles of different sizes, and other wall
geometries.

We have shown that the cross-streamline particle motion is driven by
the wall reflection of the scattered flow produced by the particles.
Namely, the reflection of the flow produced by one of the spheres
pushes the other approaching sphere across the streamlines of the
external flow towards the fluid region moving in the opposite
direction.  Due to this mutual interaction, the particles turn around
and return to infinity without passing each other -- they exchange
their vertical positions instead.

The significance of our results stems from the fact that the
wall-induced transverse particle displacements associated with
position-swapping binary encounters constitute the sole mechanism for
cross-streamline migration of spherical non-Brownian particles at low
suspension concentrations (if there are no non-hydrodynamic forces).
Such migration was indeed observed by \cite{Zarraga-Leighton:2002} who
found anomalously large self-diffusion coefficient in a confined
suspension of spheres.  Up till now, however, their measurements have
remained unexplained.  In this paper we show that the swapping
mechanism quantitatively explains the unexpected results of their
experiments.

A sequence of uncorrelated binary position-swapping particle
encounters in a monodisperse suspension causes a macroscopic
self-diffusion process.  In multi-component systems a sequence of such
encounters results in mutual diffusivity.  Moreover, the swapping
trajectories produce migration not only in the velocity-gradient
direction (considered herein) but also in the direction of fluid
vorticity.

In addition to our two-particle results, we have found that in the
wall presence there is a domain of recirculating streamlines around a
single sphere in shear flow.  Such recirculating streamlines result in
fluid mixing in suspension flows, so this phenomenon may have
important microfluidic applications.  Furthermore, the trajectory
reversal associated with particle swapping behavior may prevent
near-contact particle encounters by causing the particles to separate
before arriving into a near-contact configuration.  Such a
hydrodynamic shielding effect may be especially important for a chain
of particles with slightly different cross-streamline positions in a
microfluidic channel. Thus, proper understanding of the
position-swapping mechanism may contribute towards finding better
methods of controlling particle motions in microfluidic devices.

Other consequences of the particle-swapping mechanism will be described in our
forthcoming publications.  In particular, we will show that binary particle
collisions in a dilute suspension under shear may produce a layered particle
distribution.  Our preliminary results concerning such a layering process have
already been presented \citep*{Zurita_Gotor-Blawzdziewicz-Wajnryb:2005-C}.

We would like to thank our reviewers for calling our attention to the
experimental results by \cite{Zarraga-Leighton:2001}, and for
providing some alternative explanations of the swapping mechanism.  We
would also like to acknowledge helpful suggestions by David Leighton
regarding explanation of the swapping mechanism in terms of the wall
reflection of the stresslet, using equation \refeq{self-diffusivity
swap -- non-dimensional integral} to evaluate the self-diffusion
coefficient, and showing the asymptotic relation \refeq{scaling of
diffusivity with gap}.  This work was supported by NSF grant
CTS-0348175 (MZG and EW), NASA grant NAG3-2704 (MZG), Junta de
Andalucia grant EXC/2005/TEP-985, and by Polish Ministry of Science
and Higher Education grant N501 020 32/1994 (EW).

\appendix

\section{Flow reflected from rigid wall}
\label{Flow reflected from rigid wall}

According to the \cite{Lorentz:1907} expression, the wall reflection
$\scatteredFlow^*$ of the flow field \refeq{flow scattered by sphere
in unbounded shear flow} in a rigid wall at $z=0$ can be expressed by
the following formulas \citep{Bhattacharya-Blawzdziewicz:2002},
\begin{equation}
\label{Lorentz reflection}
\scatteredFlow^*=\oP(\hat{\bf R}_0+z_1\hat{\bf R}_1
  +z_1^2\hat{\bf R}_2)\cdot\scatteredFlow
\end{equation}
where
\begin{equation}
\label{R_0}
\hat{\bf R}_0=-{\bf I}_z-2\zField\bnabla\ez+\zField^2\nabla^2{\bf I},
\end{equation}
\begin{equation}
\label{R_1}
\hat{\bf R}_1=-2\bnabla\ez+2\zField\nabla^2{\bf I},
\end{equation}
\begin{equation}
\label{R_2}
\hat{\bf R}_2=\nabla^2{\bf I}.
\end{equation}
In the above relations, $\zField=z-z_1$ is the coordinate relative to
the source position, ${\bf I}$ is the identity tensor, ${\bf I}_z={\bf
I}-2\ez\ez$, and $\oP$ is the reflection operator with respect to the
plane $z=0$, 
\begin{equation}
\label{inversion operator}
[\oP{\bf w}](x,y,z)={\bf I}_z\cdot{\bf w}(x,y,-z).
\end{equation}
By inserting \refeq{flow scattered by sphere in unbounded shear flow}
into \refeq{Lorentz reflection} and evaluating the order of magnitude
of different terms, we find that the overall decay of the reflected
flow $\scatteredFlow^*$ with the wall-particle distance $z_1$ is
$O((d/z_1)^2)$.  Equations \refeq{flow scattered by sphere in
unbounded shear flow} and \refeq{Lorentz reflection} also indicate
that the vertical component $\scatteredFlowComponent{z}^*$ is an odd
function of the variable $x$.  Therefore, for $x/z_1\ll1$ we find
\begin{equation}
\label{scaling of flow reflected from rigid wall}
\scatteredFlowComponent{z}^*\sim\left(\frac{d}{z_1}\right)^2\frac{x}{z_1}.
\end{equation}

For a given polar angle $\phi$, equation \refeq{leading-order behavior
of image flow} is obtained from \refeq{scaling of flow reflected from
rigid wall} and an assumption that $z_1\sim H$. To obtain relation
\refeq{scaling of Delta crit}, we use \refeq{flow scattered by sphere
in unbounded shear flow}, \refeq{leading-order behavior of image
flow}, and the observation that the dependence of both
$\scatteredFlowComponent{z}$ and $\scatteredFlowComponent{z}^*$ on the
polar angle $\phi$ is $\cos\phi$ (which follows from symmetry, as
explained in Appendix \ref{New appendix}).

\section{Location of zeros of $\relativeVelocity_z$ in the plane $\zOffset=0$}
\label{New appendix}

The critical interparticle distance $\rhoOffsetCrit$ does
not depend on the orientation of the particle pair in the plane
$\zOffset=0$, which can be shown by decomposing the external flow
\refeq{external velocity} into the longitudinal component (along the
line connecting particle centers) and transverse component (normal to
this line),
\begin{equation}
\label{flow decomposition}
\externalFlow=\externalFlow_\parallel+\externalFlow_\perp.
\end{equation}
By symmetry, the transverse velocity $\externalFlow_\perp$ does not
produce any vertical particle motion.  Therefore, only the
longitudinal problem defines the location of zeroes of
$\relativeVelocity_z$.  Moreover, the location of points
$\relativeVelocity_z=0$ does not vary with the magnitude of the
projection $\externalFlow_\parallel$.  Thus the distance
$\rhoOffsetCrit$ is independent of the orientation of the vector
$\rhoOffsetVector$.

By a similar argument one can demonstrate that the upstream
swapping-trajectory domain in the variables ($\yOffset,\zOffset$) is
delimited by a straight line section $\zOffset=\const$ for
$\yOffset\le\yOffset_\crit$ (as shown in the top panel of figure
\ref{boundary lines A}).  The straight-line section corresponds to
trajectories that pass through the circle of zeros of
$\relativeVelocity_z$.  Since the transverse component
$\externalFlow_\perp$ of the external flow does not produce any
vertical particle motion, for all initial orientations of the particle
pair on the circle $\rhoOffset=\rhoOffsetCrit$ the vertical evolution
of the particles is thus the same, except for a varying time scale.

Using the flow decomposition \refeq{flow decomposition} one can also
show that for a given $\rhoOffset$, the vertical velocity
$\relativeVelocity_z$ varies as $\cos\phi$ with the polar angle $\phi$
of the relative-position vector $\rhoOffsetVector$
($\relativeVelocity_z$ varies with $\phi$ in the same way as the flow
component $\externalFlow_\parallel)$.  The plots of
$\relativeVelocity_z$ vs $\xOffset$ shown in figure \ref{signo vz} are
thus sufficient to determine $\relativeVelocity_z$ for all
orientations of a particle pair in a given plane $z=\const$.

We would also like to mention that our numerical results indicate that
additional changes of sign in the relative vertical velocity
$\relativeVelocity_z$ may occur at large interparticle distances.
This would lead to more complex particle recirculation patterns than
those described herein.  However, the magnitude of $\Delta U_z$ at
large distances is extremely small due to the exponential decay of the
vertical velocity \citep{Bhattacharya-Blawzdziewicz-Wajnryb:2006}.
Therefore, domains of trajectories with different topology are limited
to very small regions in the configurational space. We note that
multiple changes in sign of the velocity field in a parallel-wall
channel were described by \cite{Hackborn:1990} for a flow produced by
a vertical rotlet.

\section{Proof of relation \refeq{z Offset for large H}}
\label{Appendix}

As we have shown in our previous publications
\citep{%
Bhattacharya-Blawzdziewicz-Wajnryb:2006,%
Bhattacharya-Blawzdziewicz-Wajnryb:2006a%
}, 
the vertical velocity components for two spheres in Stokes flow in a
parallel-wall channel decay exponentially on the lengthscale $H$.  Thus, the
vertical offset of the particles along the pair trajectory evolves only for
$\xOffset\lesssim H$.  The relative lateral velocity of the particles in shear
flow \refeq{external velocity} is $\relativeVelocity_x\sim\shearRate\zOffset$,
where $\zOffset$ is a typical vertical offset along the trajectory.
Therefore,
\begin{equation}
\label{time scale for vertical evolution}
t_0\sim\shearRate^{-1}H/\zOffset
\end{equation}
is the timescale for the evolution of the vertical particle positions.  During
the time interval \refeq{time scale for vertical evolution} the vertical
offset $\zOffset$ may attain the magnitude
\begin{equation}
\label{magnitude of z offset}
\zOffset\sim t_0\relativeVelocity_z,
\end{equation}
where $\relativeVelocity_z$ is the typical value of the vertical component of
the relative particle velocity.

According to equation \refeq{flow scattered by sphere in unbounded
shear flow}, the perturbation velocity field $\scatteredFlow$ produced
by a spherical particle of diameter $d$ in an unbounded shear flow
with the shear rate $\shearRate$ scales as
$\scatteredFlow\sim\shearRate d(d/\rField)^2$, where $\rField$ is the
distance from the particle.  In a wall-bounded system, this
perturbation flow interacts with the walls at the distance
$\rField\sim H$ and then acts on the second particle at a distance
$\Delta x \lesssim H$, producing its vertical motion.  Therefore, the
relative vertical velocity of the particles has the magnitude
\begin{equation}
\label{magnitude of z velocity}
\relativeVelocity_z\sim\shearRate d(d/H)^2.
\end{equation}
Relation \refeq{z Offset for large H} is obtained by combining equations
\refeq{time scale for vertical evolution}--\refeq{magnitude of z velocity}.


\end{document}